\documentclass[manuscript]{aastex}

\begin{document}

\title{New Astrophysical Reaction Rate for the  $^{12}\textrm{C}(\alpha,\gamma)^{16}\textrm{O}$ Reaction}

\author{Zhen-Dong An\altaffilmark{1,2}, Yu-Gang Ma\altaffilmark{1,3}, Gong-Tao Fan\altaffilmark{1}, Yong-Jiang Li\altaffilmark{1}, Zhen-Peng Chen\altaffilmark{4}, and Ye-Ying Sun\altaffilmark{5}}

\affil{1. Shanghai Institute of Applied Physics, Chinese Academy of Sciences, Shanghai 201800, China}
\affil{2. University of Chinese Academy of Sciences, Beijing 100049, China}
\affil{3. ShanghaiTech University, Shanghai 200031, China}
\affil{4. Department of Physics, Tsinghua University, Beijing 100084, China }
\affil{5. Department of Materials, Tsinghua University, Beijing 100084, China}

\email{Email: ygma@sinap.ac.cn}
\email{Email: zhpchen@tsinghua.edu.cn}

\begin{abstract}
A new astrophysical reaction rate for $^{12}$C($\alpha,\gamma$)$^{16}$O has been evaluated
on the basis of a global R-matrix fitting to the available experimental data. The reaction rates
of $^{12}$C($\alpha,\gamma$)$^{16}$O for stellar temperatures between 0.04 $\leq$ $T_9$ $\leq$ 10
are provided in a tabular form and by an analytical fitting expression. At $T_9$ = 0.2,
the reaction rate is (7.83 $\pm$ 0.35)$\times10^{15}$ $\rm{cm^3 mol^{-1} s^{-1}}$,
where stellar helium burning occurs.
\end{abstract}


\section{Introduction}

Astrophysical reaction rates are of great importance in studies of the stellar nucleosynthesis
and the stellar evolution. During stellar helium burning the rates of 3$\alpha$ and
{$^{12}$C($\alpha,\gamma$)$^{16}$O} reaction, in competition with one another, determine the time
scale of this phase and the relative abundances of $^{12}$C and $^{16}$O in a massive star.
The reaction rate of 3$\alpha$ process is known to have an uncertainty about 10\%~\citep{fyn05},
at the astrophysical temperatures (0.2$\times$$10^{9}$ K); while such accuracy is not the case
for {$^{12}$C($\alpha,\gamma$)$^{16}$O}reaction yet relevant for the rise time of the type I
supernova light curves~\citep{dom01}, the production of important radioactive nuclei $^{26}$Al,
$^{44}$Ti, and $^{60}$Fe~\citep{tur10}, the size and mass of Fe core for a pre-supernova
star~\citep{woo03}, and the formation of X-ray black hole binaries~\citep{bro01} and neutron
star~\citep{bro98,wen13} in massive stars.

Experimental investigations on the reaction rate $N_{\rm A}\langle\sigma \emph{v}\rangle$ are calculated
with the following standard formula~\citep{rol88},
\begin{equation}\label{rate}
   N_{\rm A}\langle\sigma \emph{v}\rangle=\Big(\frac{8}{\pi\mu}\Big)^{1/2}\frac{N_{\rm A}}{{(k_{\rm B}T)}^{3/2}}
    \times\int_{0}^{\infty}dES(E)\exp(-\sqrt{\frac{E_{\rm G}}{E}}-\frac{E}{k_{\rm B}T}),
\end{equation}
where $N_{\rm A}$ refers to the Avogadro's constant, $\mu$ is the reduced mass of entrance channel
$^{12}$C+$\rm \alpha$, $k_{\rm B}$ is the Boltzmann constant, and
$E_{\rm G}={(2\pi\alpha Z_{\rm \alpha}Z_{\rm C})}^2\mu c^2/2$ is the Gamow energy with
the fine-structure constant $\rm \alpha$. The function S(E)= $\sigma(E) E \exp(2\pi\eta)$ is the total
S factor of {$^{12}$C($\alpha,\gamma$)$^{16}$O}, where $\eta$ = $Z_{\rm \alpha}Z_{\rm C}e^2/(\hbar \nu)$
is the Sommerfeld parameter, and $\sigma(E)$ is the cross section. For each temperature of $T_9$
(temperature in units of 10$^9$ K), the rate is obtained by Eq.(1) with corresponding data of the
S(E) factor.

The difficulty in measuring the S factor of {$^{12}$C($\alpha,\gamma$)$^{16}$O} reaction results from
the extremely small $\sigma$($E_{0}$), which is about $10^{-17}$ b at 0.3 MeV, where the helium burning
occurs. The observed S(E) factors are focused on the energy region of $E_{\rm {c.m.}}>$ 0.9 MeV, which
means that an extrapolation cannot be evaded at the present. It remains a challenging task to obtain
the S(E) factor for the {$^{12}$C($\alpha,\gamma$)$^{16}$O} reaction in part due to the complicated
level structure of $^{16}$O nucleus~\citep{deb13,Ma14}.

The {$^{12}$C($\alpha,\gamma$)$^{16}$O} reaction rates at astrophysical temperatures are dominated
by resonances states in the compound nucleus $^{16}$O. The rates based upon the different extrapolation
and fitting models to the parts of existing S(E) factors measurements, such as the potential models
and $R$-matrix (or $K$-matrix) theory, were reported by several research teams. Representative results
for the rates from $R$-matrix (or $K$-matrix) theory were provided by \cite{cau88} (hereafter
CF88),~\cite{buc96a},~\cite{ang99} (hereafter NACRE), and \cite{kun02}. Two recent compilations,
\cite{kat12} and~\cite{xu13} (hereafter NACRE II), are mainly based on the potential models.
However, for the corresponding reaction rate at $T_9$ = 0.2 of these compilations, the published
S-factor at 0.3 MeV disagrees at the 10\% level (see Table 1 of~\cite{an15}) with the quoted
uncertainties about twice as large as estimated for precision modeling efforts~\citep{woo07}.

In~\cite{an15}, we report a reduced $R$-matrix theory to make the global fitting to
plenty of complementary experimental data about the $^{16}$O compound nucleus. These complementary data
effectively help us to understand the concrete effect of the $^{16}$O nucleus for the S(E) factor
and the reaction rate. Based on the published S factor estimates, the updated astrophysical reaction rates
of {$^{12}$C($\alpha,\gamma$)$^{16}$O} are presented and compared with the previous published
reaction rates in this paper.

\section{Reaction Rates}

\subsection{The uncertainty of S-factors}

The S factors of $^{12}$C($\alpha,\gamma$)$^{16}$O is constituted by several resonant peaks with
strong interference pattern. And the complicated mechanism of this reaction results in unpredictable
interference effects from the first principles \citep{kun02}. The global fitting for $^{16}$O system
with a multilevel, multichannel $R$-matrix allows simultaneous analysis of differential cross section
data and angle-integrated cross section of $^{16}$O compound nucleus~\citep{an15}.
Multichannel $R$-matrix analysis provides the possibility of reducing the uncertainties in the
extrapolated total and partial S factors of ${}^{12}$C($\alpha$, $\gamma$)${}^{16}$O reaction,
and the interpretation of interference mechanism via additional constraint offered by the
simultaneous analysis of multiple reaction channels~\citep{azu10}.

The error propagation formulae~\citep{smi91} are adopted to determine the uncertainty of the S factor
in the whole energy region. Our extrapolation value is S${}_{\rm tot}$(0.3 MeV) = 162.7$\pm$7.3 keV b,
which is composed of S${}_{E10}$(0.3 MeV) = 98.0$\pm$7.0 keV b, S${}_{E20}$(0.3 MeV) = 56.0$\pm$4.1 keV b
ground state captures and of cascade captures, S${}_{\rm casc}$(0.3 MeV) = 8.7$\pm$1.8 keV b. And the
cascade transition of S$_{6.05}$ and S$_{6.13}$ at 0.3 MeV are 4.91 $\pm$ 1.11 keV.b and 0.16 $\pm$ 0.26
keV b, respectively. They are quite consistent with the constructive interference result of
S$_{6.05}$(0.3 MeV) = 4.36 $\pm$ 0.45 keV.b and destructive interference of
S$_{6.13}$(0.3 MeV) = 0.12 $\pm$ 0.04 keV b of~\cite{avi15}, which constrained contribution of
the values by measuring the asymptotic normalization coefficients (ANCs) for these states using
the $\alpha$-transfer reaction $^{6}$Li($^{12}$C, d)$^{16}$O. We adopt a similar approach for
the fitting of cascade transitions. This is one of the reasons that uncertainty of the extrapolated
S factor reduces dramatically. The values of other two cascade transitions are
S$_{7.12}$(0.3 MeV) = 0.63 $\pm$ 0.22 keV b and S$_{6.92}$(0.3 MeV) = 3.00 $\pm$ 0.42 keV b,
which are in excellent agreement with recent results of \cite{sch12}, respectively.

\subsection{New reaction rate for {$^{12}$C($\alpha,\gamma$)$^{16}$O}}

The absolute values of reaction rates, $N_{\rm A}\langle\sigma \emph{v}\rangle$ of
{$^{12}$C($\alpha,\gamma$)$^{16}$O} can be obtained by the Eq.(1) from the self-consistent total
S factor and its uncertainties. Table 1 lists 80 points of reaction rates in the temperature range of
{0.04 $\leq T_{9} \leq$ 10}. To be precise, the total reaction rate of {$^{12}$C($\alpha,\gamma$)$^{16}$O}
is achieved after multiplying $N_{\rm A}\langle\sigma \emph{v}\rangle$ with the probability densities
of the reaction partners and integrating over the energy interval. The uncertainties of the reaction
rates obtained from our R-Matrix model are also tabulated for the high and the low rate in Table 1.

According to the Gamow theory~\citep{rol88}, for the nonresonant cross section, the Gamow window
(significant integral interval) of each $T_9$ is selected in
[$E_{0}-\Delta E_{0}/2$, $E_{0}+\Delta E_{0}/2$], with $E_{0}$=($E_{\rm G}^{1/2}$$k_{\rm B}$T/2)$^{2/3}$
and $\Delta E_{0}$=(16$E_{0}$$k_{\rm B}$T/3)$^{1/2}$. Considering typical $T_9$ temperatures involved,
we have $E_{0}$ = 0.3 MeV of helium-burning start at $T_9$ = 0.2. And for the S factor data,
the chief center-of-mass energy range is 0.1-0.5 MeV. As $E_{0}$ and $\Delta E_{0}$ increase with
stellar temperature, the S(E) factor data in the higher energy range play a leading role gradually
for the reaction rate.

To understand influence of the S(E) factor of {$^{12}$C($\alpha,\gamma$)$^{16}$O} on the reaction
rate at different temperatures, probability density functions of the total reaction rates
at $T_9$ = 0.2, 1.0, 2.0, 3.0, 4.0, 5.0, 6.0, 8.0 and 10.0 are shown
in Fig.1. At $T_9$ = 0.2 and 1.0, the probability density functions are located almost
in the extrapolated S factor, without resonance peaks, so the Gamow window can be well approximated
by a Gaussian-distribution with the most effective energy $E_0$ = 0.3 MeV and $E_0$ = 0.9 MeV.
As the $T_9$ increases from 1 to 10, however, the influence of resonances of S(E) factor becomes
more and more remarkable, and the probability density functions can no longer be approximated
by the Gaussian-distribution. So, {$^{12}$C($\alpha,\gamma$)$^{16}$O} reaction rate at these
temperatures can be obtained just by the S factor measurements at energies as wide as possible.

The S${}_{\rm tot}$ measurements of \cite{sch05,sch11}, in reverse kinematics using the recoil mass
separator allowed to acquire data with a high degree of accuracy ($<$3\%) in a wide energy scope of
$E_{\rm {c.m.}}$ = 1.5 - 4.9 MeV. These data would make good restriction to probability density functions
of $T_9$ = 2.0, 3.0, and 4.0 (Fig. 1C-1E), and uncertainties of {$^{12}$C($\alpha,\gamma$)$^{16}$O}
reaction rate are smaller than 3\% from {2.0 $\leq T_{9} \leq$ 4.0}.

\cite{kun02} studied on the S factor to the ground state of $^{12}$C($\alpha,\gamma$)$^{16}$O at
higher energies, covering the $1^-_3$ ($E_{c.m.}$= 5.28 MeV) and $1^-_4$ ($E_{c.m.}$= 5.93 MeV),
using resonance parameters of \cite{til93} in the calculation. The published data in two
independent experiments ~\citep{oph76, bro73} of the ground-state transition were neglected.
Possible interference effects were included in the calculation of S${}_{E10}$
and S${}_{E20}$ by applying the $R$-matrix fitting procedures, but they were somewhat speculative,
and the results of S${}_{g.s}$ were about 2$\sim$5 times away from experimental data. Therefore,
from the probability density functions of $T_9$ = 5.0, 6.0, 8.0 and 10.0 (Fig. 1F-1I),
it can be inferred that the rate calculation of \cite{kun02} is significantly higher.

The S${}_{\rm tot}$ can be indicated according to different types of $J^\pi$ ($0^+$, $1^-$, $2^+$,
$3^-$, and $4^+$). Fig. 2(A) shows the fractional contributions of different values of $J^\pi$
to the total reaction rates in {0.04 $\leq T_{9} \leq$ 10.0}. It can be seen that $J^\pi$ = $1^-$
and $2^+$ dominate the reaction rate up to $T_9$ = 2.0. From the probability density functions of
$T_9$ = 0.2, 1.0, and 2.0 (Fig. 1A-1C), we note that the contribution stems mainly from
$J^\pi$ = $1^-$ ($E_{c.m.}$= 2.42 MeV) and $J^\pi$ = $2^+$ ($E_{c.m.}$= 2.58 MeV) levels in
$^{16}$O. The rate above $T_9$ = 2.0, the fraction from $3^-$ gradually increases with temperature.
This comes from the contribution of $J^\pi$ = $3^-$ ($E_{c.m.}$= 4.44, 5.97, and 6.10 MeV) resonances.
The contribution of $J^\pi$ = $4^+$ increases with $T_9$ first, and then decreases, having two $4^+$
resonances just at $E_{c.m.}$= 3.20 and 3.93 MeV in integral interval. Thus, the fractional
contributions of different values of $J^\pi$ reconfirmed the validity of the probability density
functions in Fig. 1.

The contributions of ground state capture and cascade captures to the reaction rate of
{$^{12}$C($\alpha,\gamma$)$^{16}$O} are illustrated in Fig. 2(B). Ground state capture
(S$_{E10}$ and S$_{E20}$) dominates the reaction rate up to $T_9$ = 0.1. And the contributions
from the cascade transitions increase with $T_9$. Beside at the important He-burning temperature
$T_9$ = 0.2, the rate is still important all the way up to $T_9$ = 5.0, because
the inverse reaction of {$^{12}$C($\alpha,\gamma$)$^{16}$O} plays an important role
in silicon burning~\citep{woo13}. So cascade transition is necessary to the precise calculation
of reaction rate.

\subsection{Comparison to Other $^{12}$C($\alpha,\gamma$)$^{16}$O Determinations}

Fig. 3 shows comparisons between our new reaction rate and previous estimates. In each panel
the dashed line shows the ratio of a previous determination to our new rate. The grey bands are
the uncertainty of the published rate, e.g., in Fig. 3(A) the edges of grey zone are reaction rate
ratios of NARCE II's limits to the principal value of our rates. The blue bands estimate
the uncertainty of our rate. Below $T_{9} \approx$ 4.0, our recommended results are within
the uncertainties of \cite{buc96a}, \cite{kun02} and NACRE II. Above $T_{9}$ = 4, the results agree
with the analysis of NACRE.

By comparing all the published rates with our present results, the principal values of each rate
are shown in Fig. 4. At astrophysical temperature of $T_{9}$ =0.2 the new rate is
about 10\% larger than the rate of NACRE II (S${}_{\rm tot}$(0.3 MeV)=148$\pm$27 keV b) and
\cite{buc96a} (S${}_{\rm tot}$(0.3 MeV)=146 keV b), about 16\% lower than the rate
of the NACRE (S${}_{\rm tot}$(0.3 MeV)=199$\pm$64 keV b), and it is quite consistent
with the adopted value of \cite{kun02} (S${}_{\rm tot}$(0.3 MeV)=165$\pm$50 keV b).

In the intermediate range {0.5 $\leq T_{9} \leq$ 3}, our recommended rate is in good agreement with
NACRE II. The temperature dependence of our recommended value differs significantly from
the rates of \cite{kat12}, which stems from the higher total S factor at $1^{-}_{2}$
($E_{\rm{c.m.}}$= 2.42 MeV) resonance-peak, overestimating the cross-section of \cite{sch05}
in their calculations \citep{kat08}. In the same temperature range, the deviation from \cite{kun02}
mainly originates from the lower calculation values of total S factor from $E_{\rm{c.m.}}$= 0.5 MeV
to $E_{\rm{c.m.}}$= 2.0 MeV. The lower value of NACRE is a direct consequence of the
considered cascade transitions for the total S-factors.

For the rates above $T_{9}$ = 3, our reaction rate increases with $T_{9}$ but has lower values
than \cite{kun02} and NACRE II, because the high-energy data covering the $1^{-}_{3}$
and $1^{-}_{4}$ resonance are overestimated apparently in their calculations.

\subsection{Analytical formula}

A common form of reaction rate is an analytical formula with an appropriate parametrization
for applications in stellar models. Eq. (2) is a usual expression \citep{buc96a,kun02}.

\begin{displaymath}
    N_{\rm A}\langle\sigma \emph{v}\rangle^{AF} = \frac{a_0}{{T^2_9}(1+{a_1}T^{-2/3}_9)^2}\exp\Big[-\frac{a_2}{T^{1/3}_9}-\Big(\frac{T_9}{a_3}\Big)^{2}\Big]
    +\frac{a_4}{{T^2_9}(1+{a_5}T^{-2/3}_9)^2}\exp\Big(-\frac{a_6}{T^{1/3}_9}\Big)
\end{displaymath}
\begin{equation}\label{ratefit1}
    +\frac{a_7}{T^{3/2}_9}\exp\Big(-\frac{a_8}{T_9}\Big)
    +\frac{a_9}{T^{2/3}_9}{(1+{a_{10}}T^{1/3}_9)}\exp\Big(-\frac{a_{11}}{T^{1/3}_9}\Big).
\end{equation}

The difference between the fitting formula to tabulated rate is shown in Fig.3 (F).
The blue bands indicate the uncertainty of adopted rate in Table 1. The dotted line shows the ratio
of the adopted values of the analytical expression normalized to the adopted tabulated ones.
It is applicable in temperature range of $0.04 \leq T_9 \leq 10$ with a maximum deviation of 4.0\%
to the recommended rate in Table 1. For the most important range of $T_{9}$ = 0.1-0.3 the maximal
deviation is 1\%. And the parameters $a_0 - a_{11}$ are
$a_{0} $ = $ 4.70\times10^{8} $;
$a_{1} $ = 0.312;
$a_{2} $ = 31.8;
$a_{3} $ = 400;
$a_{4} $ = $ 1.08\times10^{15}$;
$a_{5} $ = 23.6;
$a_{6} $ = 41.3;
$a_{7} $ = $ 2.49\times10^{3} $;
$a_{8} $ = 28.5;
$a_{9} $ = $ 1.19\times10^{11}$;
$a_{10}$ = -98.0;
$a_{11}$ = 36.5.

\section{Conclusions}

New $^{12}$C($\alpha,\gamma$)$^{16}$O reaction rates in the range of $0.04 \leq T_9 \leq 10$
have been estimated from recent S factor modeling. The measurements at higher energies are
analyzed in our $R$-matrix fit, which significantly reduce the uncertainty of reaction rate at
higher temperatures. A comprehensive comparison is done between our results and the previous data.
It should be noted that, the results are obtained by theoretical extrapolation of existing
experimental data of $^{16}$O system. There could be some important factors that our model
does not include. Additional experiments and theoretical work are needed to further validate
existing expressions for the $^{12}$C($\alpha,\gamma$)$^{16}$O rate.

From the Figs. 1 and 2, it is suggested that an improved investigation of S$_{\rm casc}$
and S$_{\rm TOT}$ at $1^-_3$ and $1^-_4$ resonances, may help to further reduce the uncertainties
of reaction rates at higher temperatures. Moreover, the asymptotic normalization coefficients (ANCs)
of the corresponding states do not consist with each other in the transfer reaction
\citep{bru99,bel07,oul12,avi15} with big uncertainty, which remain to be solved. Finally,
the extrapolated S${}_{\rm tot}$(0.3 MeV) is quite sensitive to the data as close as possible
to the Gamow window. The reverse reaction ($\gamma$, $\alpha$) using a high photon flux $\gamma$-ray beam,
such as the High Intensity $\gamma$-ray Source at TUNL ~\citep{Gai12,dig15} and the under
construction of Shanghai Laser Electron Gamma Source facility in our team~\citep{xu07,luo11},
would be desirable to allow a measurement of cross sections in the pb region.

\acknowledgments

The authors would like to thank Prof. Stan Woosley and Prof. Alexander Heger for helpful discussions
of reaction rates. This work is supported partially by the National Natural Science Foundation of
China under Grant Nos. 11175233, 91126017 and 11421505 and the 973 project under contract no. 2014CB845401.

\clearpage

\begin{figure}
\epsscale{1.0}
\plotone{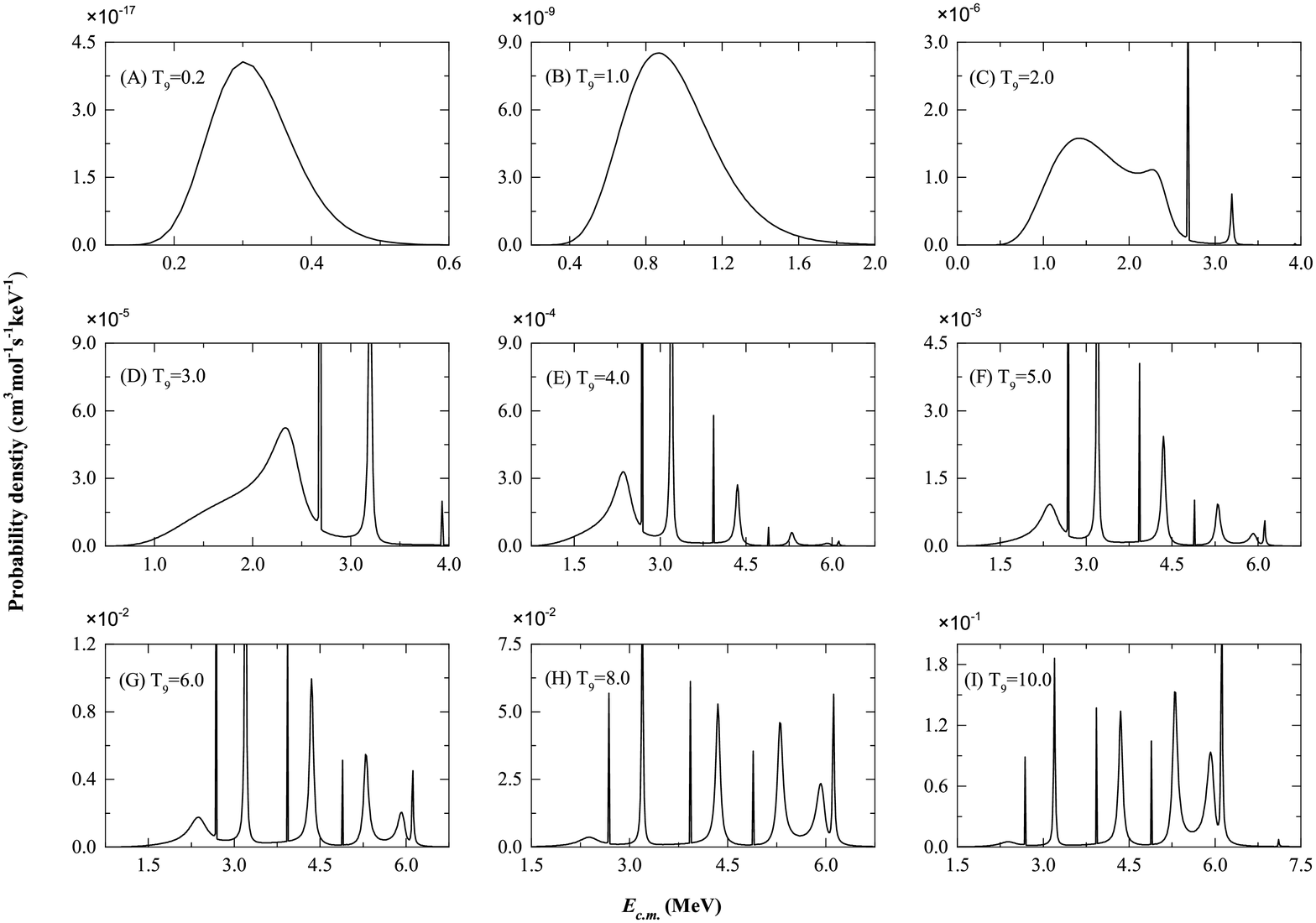}
\caption{
    Reaction rate probability density functions for $^{12}$C($\alpha,\gamma$)$^{16}$O
    at different values of $T_9$.\label{fig1}}
\end{figure}

\clearpage

\begin{figure}
\epsscale{0.55}
\plotone{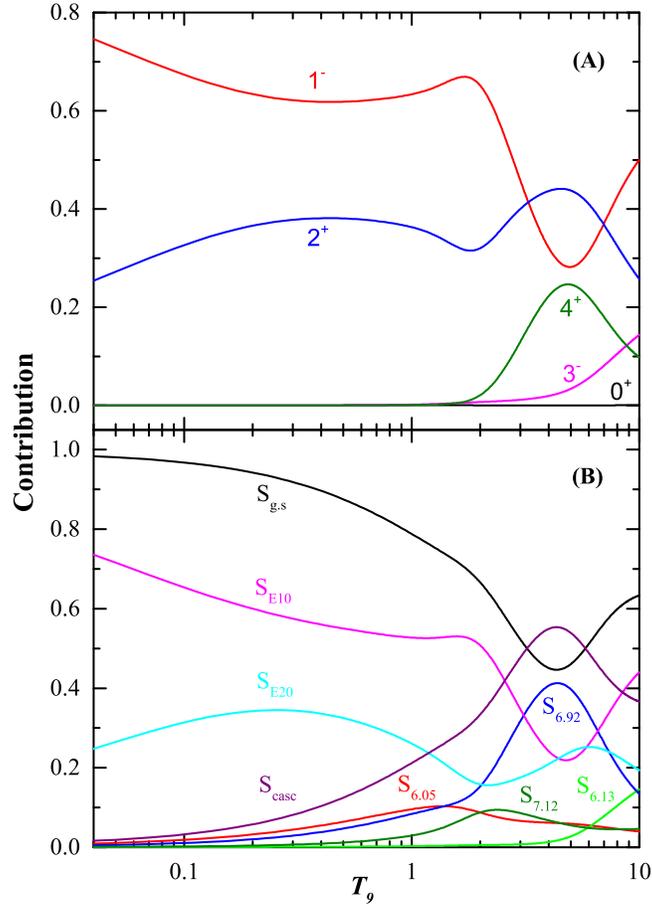}
\caption{
    Top: fractional contributions of different $J^\pi$ to the total reaction rates
    of {$^{12}$C($\alpha,\gamma$)$^{16}$O}.
    Bottom: fractional contributions of S${}_{g.s.}$ (including E${}_{10}$ and E${}_{20}$
    to the ground state) and the cascade transitions to the total rates.\label{fig2}}
\end{figure}

\clearpage
\begin{figure}
\epsscale{1.0}
\plotone{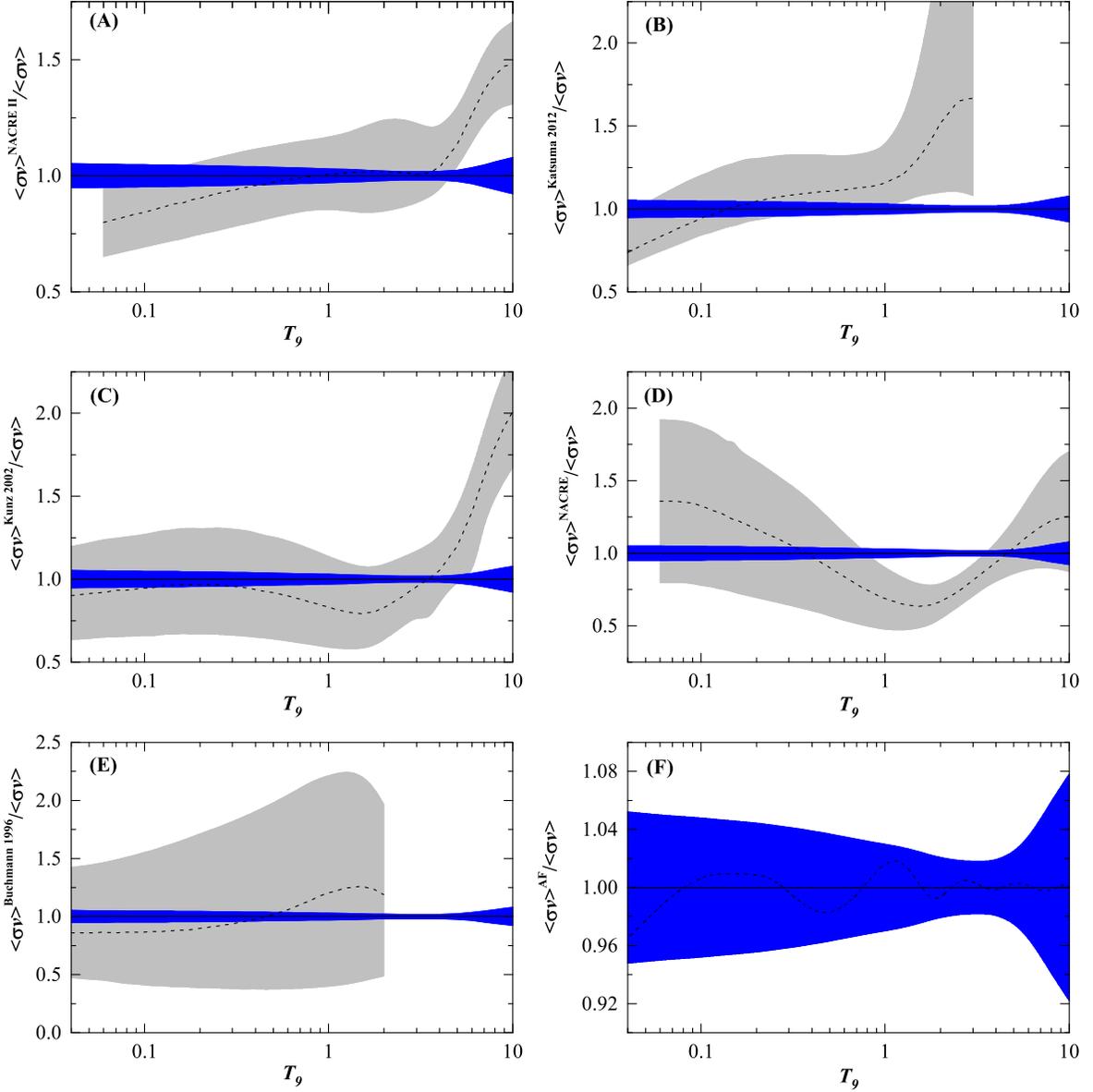}
\caption{
    Comparisons (ratio) of $^{12}$C($\alpha,\gamma$)$^{16}$O reaction rates from the compilations of
    (A) NACRE II, (B) Katsuma (2012), (C) Kunz et al. 2002, (D) NACRE and (E) Buchmann (1996)
    with our new recommended rate. Accuracy of the analytic formula according to Eq.(2)
    is shown in Fig.3(F).
    \label{fig3}}
\end{figure}

\begin{figure}
\epsscale{.70}
\plotone{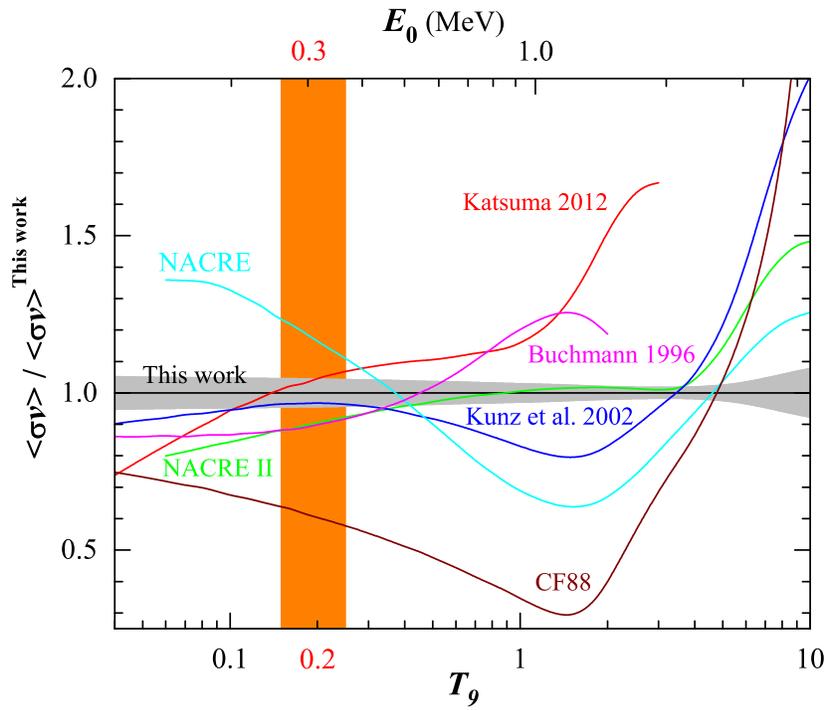}
\caption{
    Comparisons of astrophysical reaction rate of $^{12}$C($\alpha,\gamma$)$^{16}$O (including CF88)
    normalized to our new recommended rate.\label{fig4}}
\end{figure}

\begin{deluxetable}{lrrrrcrrrrr}
\tablewidth{0pt}
\tablecaption{The {$^{12}$C($\alpha,\gamma$)$^{16}$O} reaction rates (in $\rm{cm^3 mol^{-1} s^{-1}}$).
                \label{tbl-rate}}
\tablehead{
  \colhead{$T_9$}  & \colhead{$N_{\rm A}\langle\sigma \emph{v}\rangle$}  & \colhead{High}
& \colhead{Low} & \colhead{$10^n$} & \colhead{} &
  \colhead{$T_9$}  & \colhead{$N_{\rm A}\langle\sigma \emph{v}\rangle$}  & \colhead{High}
& \colhead{Low} & \colhead{$10^n$} }
\startdata
0.040  &	9.27  &	9.75  &	8.79  &	-31	& 	& 0.60   &	3.28  &	3.40  &	3.17  &	-8	\\
0.042  &	3.94  &	4.14  &	3.74  &	-30	& 	& 0.65   &	8.00  &	8.27  &	7.73  &	-8	\\
0.045  &	2.92  &	3.07  &	2.77  &	-29	& 	& 0.70   &	1.78  &	1.84  &	1.73  &	-7	\\
0.050  &	5.69  &	5.98  &	5.40  &	-28	& 	& 0.75   &	3.70  &	3.82  &	3.58  &	-7	\\
0.055  &	7.60  &	7.98  &	7.21  &	-27	& 	& 0.80   &	7.19  &	7.42  &	6.96  &	-7	\\
0.060  &	7.51  &	7.88  &	7.13  &	-26	& 	& 0.85   &	1.32  &	1.37  &	1.28  &	-6	\\
0.065  &	5.81  &	6.10  &	5.52  &	-25	& 	& 0.90   &	2.33  &	2.40  &	2.26  &	-6	\\
0.070  &	3.67  &	3.85  &	3.49  &	-24	& 	& 0.95   &	3.93  &	4.05  &	3.81  &	-6	\\
0.075  &	1.95  &	2.05  &	1.86  &	-23	& 	& 1.00   &	6.41  &	6.60  &	6.22  &	-6	\\
0.080  &	9.00  &	9.44  &	8.57  &	-23	& 	& 1.10   &	1.56  &	1.60  &	1.51  &	-5	\\
0.085  &	3.66  &	3.84  &	3.49  &	-22	& 	& 1.20   &	3.42  &	3.52  &	3.33  &	-5	\\
0.090  &	1.34  &	1.40  &	1.27  &	-21	& 	& 1.30   &	6.96  &	7.15  &	6.78  &	-5	\\
0.095  &	4.45  &	4.67  &	4.24  &	-21	& 	& 1.40   &	1.33  &	1.36  &	1.30  &	-4	\\
0.100  &	1.36  &	1.43  &	1.30  &	-20	& 	& 1.50   &	2.41  &	2.47  &	2.35  &	-4	\\
0.105  &	3.88  &	4.06  &	3.69  &	-20	& 	& 1.60   &	4.19  &	4.29  &	4.09  &	-4	\\
0.11   &	1.03  &	1.08  &	0.98  &	-19	& 	& 1.70   &	7.02  &	7.18  &	6.85  &	-4	\\
0.12   &	6.18  &	6.48  &	5.89  &	-19	& 	& 1.80   &	1.14  &	1.16  &	1.11  &	-3	\\
0.13   &	3.06  &	3.20  &	2.91  &	-18	& 	& 1.90   &	1.80  &	1.83  &	1.76  &	-3	\\
0.14   &	1.29  &	1.35  &	1.23  &	-17	& 	& 2.00   &	2.76  &	2.82  &	2.71  &	-3	\\
0.15   &	4.75  &	4.97  &	4.53  &	-17	& 	& 2.10   &  4.15  & 4.23  & 4.07  & -3	\\
0.16   &  1.56  & 1.64  & 1.49  & -16 &   & 2.20   &  6.10  & 6.22  & 5.98  & -3 \\
0.17   &  4.67  & 4.88  & 4.46  & -16 &   & 2.30   &  8.79  & 8.96  & 8.62  & -3 \\
0.18   &  1.28  & 1.34  & 1.23  & -15 &   & 2.50   &  1.73  & 1.76  & 1.69  & -2 \\
0.19   &  3.27  & 3.42  & 3.13  & -15 &   & 2.75   &  3.66  & 3.73  & 3.60  & -2 \\
0.20   &  7.83  & 8.18  & 7.48  & -15 &   & 3.00   &  7.13  & 7.26  & 7.00  & -2 \\
0.21   &  1.77  & 1.85  & 1.69  & -14 &   & 3.25   &  1.29  & 1.32  & 1.27  & -1 \\
0.22   &  3.79  & 3.96  & 3.63  & -14 &   & 3.50   &  2.20  & 2.24  & 2.16  & -1 \\
0.24   &  1.53  & 1.59  & 1.46  & -13 &   & 3.75   &  3.58  & 3.64  & 3.51  & -1 \\
0.26   &  5.29  & 5.52  & 5.07  & -13 &   & 4.00   &  5.57  & 5.67  & 5.46  & -1 \\
0.28   &  1.62  & 1.69  & 1.55  & -12 &   & 4.25   &  8.37  & 8.54  & 8.21  & -1 \\
0.30   &  4.47  & 4.66  & 4.29  & -12 &   & 4.50   &  1.22  & 1.25  & 1.20  & 0 \\
0.32   &  1.13  & 1.18  & 1.08  & -11 &   & 5.00   &  2.42  & 2.48  & 2.36  & 0 \\
0.34   &  2.65  & 2.75  & 2.54  & -11 &   & 5.50   &  4.43  & 4.56  & 4.31  & 0 \\
0.36   &  5.81  & 6.04  & 5.57  & -11 &   & 6.00   &  7.60  & 7.87  & 7.34  & 0 \\
0.38   &  1.20  & 1.25  & 1.15  & -10 &   & 6.50   &  1.23  & 1.28  & 1.18  & 1 \\
0.40   &  2.37  & 2.46  & 2.27  & -10 &   & 7.00   &  1.90  & 1.99  & 1.81  & 1 \\
0.42   &  4.46  & 4.63  & 4.28  & -10 &   & 7.50   &  2.79  & 2.94  & 2.64  & 1 \\
0.45   &  1.07  & 1.11  & 1.03  & -9  &   & 8.00   &  3.95  & 4.18  & 3.71  & 1 \\
0.50   &  3.91  & 4.05  & 3.76  & -9  &   & 9.00   &  7.13  & 7.63  & 6.64  & 1 \\
0.55   &  1.21  & 1.25  & 1.17  & -8  &   & 10.0   &  1.15  & 1.24  & 1.06  & 2 \\
\enddata
\end{deluxetable}

\end{document}